\documentclass[twocolumn,aps,pra,amsmath,amssymb]{revtex4-1}
\usepackage[latin9]{inputenc}
\usepackage{mathrsfs}
\usepackage{amsmath}
\usepackage{amssymb}
\usepackage{graphicx}
\usepackage{esint}

\makeatletter
\usepackage{epsfig}
\usepackage{bm}
\usepackage{dcolumn}
\usepackage{color}

\makeatother

\begin{document}

\title{Resonantly interacting $p$-wave Fermi superfluid in two dimensions:
Tan's contact and breathing mode}

\author{Hui Hu}

\affiliation{Centre for Quantum and Optical Science, Swinburne University of Technology,
Melbourne, Victoria 3122, Australia}

\author{Xia-Ji Liu}

\affiliation{Centre for Quantum and Optical Science, Swinburne University of Technology,
Melbourne, Victoria 3122, Australia}

\date{\today}
\begin{abstract}
Inspired by the renewed experimental activities on $p$-wave resonantly
interacting atomic Fermi gases, we theoretically investigate some
experimental observables of such systems at zero temperature in two
dimensions, using both mean-field theory and Gaussian pair fluctuation
theory. These observables include the two $p$-wave contact parameters
and the breathing mode frequency, which can be readily measured in
current cold-atom setups with $^{40}$K and $^{6}$Li atoms. We find
that the many-body component of the two contact parameters exhibits
a pronounced peak slightly above the resonance and consequently leads
to a dip in the breathing mode frequency. In the resonance limit,
we discuss the dependence of the equation of state and the breathing
mode frequency on the dimensionless effective range of the interaction,
$k_{F}R_{p}\ll1$, where $k_{F}$ is the Fermi wavevector and $R_{p}$
is the effective range. The breathing mode frequency $\omega_{B}$
deviates from the scale-invariant prediction of $\omega_{c}=2\omega_{0}$,
where $\omega_{0}$ is the trapping frequency of the harmonic potential.
This frequency shift is caused by the necessary existence of the effective
range. In the small range limit, we predict that the mode frequency
deviation at the leading order is given by, $\delta\omega_{B}\simeq-(\omega_{0}/4)\ln^{-1}(k_{F}R_{p})$.
\end{abstract}

\maketitle

\section{Introduction}

The realization of Feshbach resonances in ultracold atoms provides
a unique opportunity to explore fascinating quantum many-body phenomena
\cite{Chin2010}. By precisely tuning the $s$-wave scattering length
using an external magnetic field, one can now routinely produce a
stable cloud of strongly interacting fermions and observe novel Fermi
superfluidity at the cusp of the crossover from a Bardeen-Cooper-Schrieffer
(BCS) superfluid to a Bose-Einstein condensate (BEC) \cite{Eagles1969,Leggett1980,Nozieres1985,SadeMelo1993,Regal2004,Zwierlein2004,Hu2006,Diener2008,Bloch2008,Giorgini2008}.
The manipulation of high-partial-wave interatomic interactions is
also possible. In particular, about fifteen years ago the experimental
demonstration of $p$-wave Feshbach resonances in $^{40}$K and $^{6}$Li
atoms \cite{Regal2003,Zhang2004,Gunter2005,Schunck2005,Gaebler2007,Fuchs2008,Inada2008}
opened the exciting perspective of creating a topological $p$-wave
Fermi superfluid, which hosts non-trivial non-Abelian excitations
at its edges or in its vortex cores - the so-called Majorana fermions
- that could enable topological quantum computation \cite{Read2000,Ivanov2001,Kitaev2003,Nayak2008}.
Unfortunately, unlike a strongly interacting Fermi gas at BEC-BCS
crossover, the $p$-wave resonantly interacting systems generally
suffers from a heating problem due to serious loss in atom number
and can hardly reach a low-temperature equilibrium state. Therefore,
there is no significant experimental progress, in spite of a lot of
interesting theoretical investigations at the early stage \cite{Gurarie2007},
exploring different aspects of a strongly interacting $p$-wave Fermi
superfluid, such as the zero-temperature phase diagram \cite{Botelho2005,Ho2005,Gurarie2005,Cheng2005,Iskin2006,Cao2013},
the superfluid transition temperature in three dimensions (3D) \cite{Ohashi2005,Inotani2012,Inotani2015}
and the Berezinskii-Kosterlitz-Thouless (BKT) phase transition in
two dimensions (2D) \cite{Cao2017}. 

This situation is much improved over the past few years \cite{Luciuk2016,Waseem2017,Yoshida2018,Wassem2018}.
After a quench in the external magnetic field to the resonance limit,
a quasi-equilibrium state of a 3D strongly interacting $p$-wave Fermi
gas has been observed \cite{Luciuk2016}, and the contact parameters,
which characterize the universal short-distance and large-momentum
behavior of the system \cite{Tan2008a,Tan2008b,Tan2008c,Yoshida2015,Yu2015,He2016,Peng2016,Zhang2017,Yao2018,Inotani2018},
have been measured using radio-frequency (rf) spectroscopy \cite{Luciuk2016}.
Most recently, the atom loss close to the resonance has been found
to reduce significantly in lower dimensions \cite{Waseem2017}, as
theoretically predicted \cite{Levinsen2008,Fedorov2017}. These experimental
advances suggest the possibility of realizing a 2D strongly interacting
$p$-wave Fermi superfluid in future experiments.

Motivated by this possibility, here we present a detailed theoretical
study of two important experimental observables of a $p$-wave Fermi
superfluid at \emph{zero} temperature: the two $p$-wave contact parameters
and the breathing mode frequency. The investigation is based on our
recent results of the zero-temperature equations of state \cite{Hu2018},
which are reliably calculated using the Gaussian pair fluctuation
(GPF) theory beyond mean-field \cite{Hu2006,Diener2008,Hu2007,He2015}.
A finite-temperature investigation is also possible, by applying the
Noziéres and Schmitt-Rink theory above the superfluid phase transition
\cite{Nozieres1985,SadeMelo1993}. We note that, at sufficient high
temperatures close to the Fermi degenerate temperature, the calculations
of $p$-wave contact parameters and breathing mode frequency were
recently performed by Yi-Cai Zhang and Shizhong Zhang \cite{Zhang2017},
using the virial expansion theory \cite{Yu2009,Liu2009,Hu2011,Liu2013}.

In this work, we are particularly interested in the breathing mode
frequency right at the resonance. In three dimensions, an $s$-wave
resonantly interacting Fermi gas acquires scale-invariant zero-energy
wave-functions, which are eigenstates of the dilation operator \cite{Werner2006}.
In the presence of an isotropic harmonic trap with frequency $\omega_{0}$,
there is a hidden symmetry $SO(2,1)$, yielding a scale-invariant
breathing mode frequency $\omega_{c}=2\omega_{0}$ \cite{Werner2006}.
In two dimensions, this hidden symmetry was nicely explained by Pitaevskii
and Rosch using the same contact $s$-wave interatomic interaction,
which is scale-invariant classically \cite{Pitaevskii1997}. However,
the quantum renormalization of the $s$-wave contact interaction necessarily
introduces a new length scale of the 2D scattering length and explicitly
breaks the scale-invariance of the interaction. Therefore, the breathing
mode frequency deviates from the classically invariant value of $\omega_{c}$,
i.e., $\delta\omega_{B}=\omega_{B}-\omega_{c}\neq0$. This frequency
shift is now referred to as quantum anomaly \cite{Hofmann2012,Taylor2012,Cao2012,Vogt2012,Mulkerin2018,Holten2018,Peppler2018,Hu2019}.
In our case of a \emph{resonantly} interacting $p$-wave interaction,
where the 2D scattering area disappears, we anticipate that the system
may also have scale-invariant zero-energy wave-functions if there
is no length scale set by interactions, and in the presence of isotropic
harmonic trap it has the scale-invariant breathing mode frequency
$\omega_{c}$. This is unfortunately not true. The renormalization
of $p$-wave interaction necessarily gives a length scale of the effective
range of interactions. The breathing mode frequency then deviates
from $\omega_{c}$. We find that this frequency shift in the $p$-wave
channel is much larger than its $s$-wave counterpart of quantum anomaly
and could be more easily measured in experiments.

The rest of the paper is set as follows. In the next section (Sec.
II), we provide the Hamiltonian of a 2D spinless Fermi gas near a
$p$-wave Feshbach resonance described by a separable interaction
potential. We show how to calculate the scattering area $a_{p}$ and
the effective range $R_{p}$ for the separable potential. As discussed
in Sec. III, this enables us to obtain the equations of state of the
system as functions of $a_{p}$ and $R_{p}$. We derive the analytic
mean-field equations and present the numerical GPF results beyond
mean-field for the equations of state. In Sec. IV, we discuss the
two $p$-wave contacts and the related breathing mode frequency. In
Sec. V, we focus on the resonance limit and discuss the significant
frequency shift from the scale-invariant frequency $\omega_{c}$,
due to the existence of the effective range. Finally, Sec. VI is devoted
to conclusions.

\section{Model Hamiltonian and two-body scattering}

A spinless 2D $p$-wave interacting Fermi gas of $N$ atoms can be
described by the Hamiltonian (with the area $A=1$) \cite{Botelho2005,Hu2018},
\begin{equation}
{\cal H}=\sum_{{\bf k}}\xi_{{\bf k}}\psi_{{\bf k}}^{\dagger}\psi_{{\bf k}}+\frac{1}{2}\sum_{{\bf k},{\bf k}^{\prime},{\bf q}}V_{\mathbf{k}\mathbf{k}'}b_{\mathbf{k}\mathbf{q}}^{\dagger}b_{\mathbf{k'}\mathbf{q}},
\end{equation}
where $\psi_{{\bf k}}^{\dagger}$ ($\psi_{{\bf k}}$) is the creation
(annihilation) field operator for atoms, $\xi_{{\bf k}}\equiv\epsilon_{{\bf k}}-\mu=\hbar^{2}\mathbf{k}^{2}/(2M)-\mu$
is the single-particle dispersion with mass $M$ and chemical potential
$\mu$, and $b_{\mathbf{k}\mathbf{q}}^{\dagger}\equiv\psi_{{\bf k}{\bf +q}/2}^{\dagger}\psi_{-{\bf k}{\bf +q}/2}^{\dagger}$
is the composite operator that creates a pair of atoms with center-of-mass
momentum $\mathbf{q}$. The inter-particle interaction takes a separable
form with the chiral $p_{x}+ip_{y}$ symmetry \cite{Nozieres1985,Botelho2005,Ho2005,Hu2018},
\begin{equation}
V_{\mathbf{k}\mathbf{k}'}=\lambda\Gamma\left({\bf k}\right)\Gamma^{*}\left({\bf k'}\right).
\end{equation}
Here, $\lambda$ is the bare interaction strength and

\begin{equation}
\Gamma\left({\bf k}\right)=\frac{\left(k/k_{F}\right)}{\left[1+\left(k/k_{0}\right)^{2n}\right]^{3/2}}e^{i\varphi_{{\bf k}}}
\end{equation}
is a dimensionless regularization function with the cut-off momentum
$k_{0}$, polar angle $\varphi_{{\bf k}}$ and exponent $n$ that
is introduced for the convenience of numerical calculations. The Fermi
wavevector $k_{F}$ is related to the number density of atoms $n_{2D}=N/A$
by the relation, $k_{F}=\sqrt{4\pi n_{2D}}$. Our choice of the chiral
$p_{x}+ip_{y}$ channel is motivated by the phase diagram established
by Gurarie et al. \cite{Gurarie2005}. Although experimentally the
Feshbach resonances for $m=0$ and $m=\pm1$ are nearly degenerate,
at low temperature the system will spontaneously break the spin-rotational
symmetry and condense into the $p_{x}+ip_{y}$ superfluid state.

We may replace $\lambda$ with a characteristic energy $E_{b}$ by
solving the following two-body problem at zero center-of-mass momentum
\cite{Botelho2005,Hu2018},
\begin{equation}
2\epsilon_{\mathbf{k}}\Psi_{\mathbf{k}}+\sum_{\mathbf{k}'}V_{\mathbf{kk}'}\Psi_{\mathbf{k}'}=E_{b}\Psi_{\mathbf{k}},
\end{equation}
where $\Psi_{\mathbf{k}}$ is the two-body wave-function in momentum
space and $\mathbf{k}$ is the relative momentum of two particles.
By using the separability of the interaction potential, after some
algebra, it is easy to find that,
\begin{equation}
\frac{1}{\lambda}=-\mathscr{P}\sum_{{\bf p}}\frac{\left|\Gamma\left({\bf p}\right)\right|^{2}}{2\epsilon_{{\bf p}}-E_{b}},\label{eq:renorm}
\end{equation}
where $\mathcal{\mathscr{P}}$ stands for taking Cauchy principal
value. As we shall see in the next subsection, $E_{b}$ is related
to the 2D scattering area $a_{p}$ (see Eq. (\ref{eq:ap}) below).
It can be either negative or positive \cite{Botelho2005,Hu2018}.
In the former case, the so-called BEC side, it is simply the ground-state
energy of a two-body bound state and the associated binding energy
$\varepsilon_{B}=-E_{b}>0$. In the latter, it may be viewed as a
scattering energy $E_{b}=\hbar^{2}\mathbf{k}_{b}^{2}/M>0$ for two
particles colliding with a characteristic relative momentum $\mathbf{k}_{b}$
within the two-particle continuum. In this case, the sum at the right-hand-side
of Eq. (\ref{eq:renorm}) is not well-defined and we have taken the
Cauchy principal value of the sum to remove possible ambiguity. From
now on, for convenience we name $E_{b}$ as the scattering energy,
in spite of the fact that it can take negative values on the BEC side.

In the previous work \cite{Hu2018}, we determined the equations of
state of the 2D $p$-wave Fermi superfluid as functions of the parameters
($E_{b},k_{0},n=1$). Here, for the purpose of calculating the $p$-wave
contacts and breathing mode frequency, it is more useful to parameterize
the inter-particle interaction by using the 2D scattering area $a_{p}$
and the effective range of interactions $R_{p}$, which are formally
defined through the $p$-wave phase shift $\delta_{p}\left(k\right)$
\cite{Zhang2017,Levinsen2008,Hu2018},
\begin{equation}
k^{2}\cot\delta_{p}\left(k\right)=-\frac{1}{a_{p}}+\frac{2k^{2}}{\pi}\ln\left(R_{p}k\right)+\cdots.\label{eq:PhaseShift}
\end{equation}
As shown in Appendix of our previous work \cite{Hu2018}, we find
that $R_{p}=k_{0}^{-1}$ in the limit of $n\rightarrow\infty$. In
the following, we derive the general expressions of $a_{p}$ and $R_{p}$
for an arbitrary exponent $n$. This is necessary, since we have to
take a finite value of $n$ in actual numerical calculations. At low
energy, all the physical results of interest should be functions of
$a_{p}$ and $R_{p}$, independent of the different choice for $n$.

\subsection{The expressions of $a_{p}$ and $R_{p}$}

To relate the bare interaction strength $\lambda$ to the scattering
parameters, we calculate the two-body $T$-matrix in vacuum \cite{Hu2018},
\begin{equation}
T\left({\bf k},{\bf k};E\right)=\left|\Gamma\left({\bf k}\right)\right|^{2}\left[\frac{1}{\lambda}+\sum_{{\bf p}}\frac{\left|\Gamma\left({\bf p}\right)\right|^{2}}{2\epsilon_{{\bf p}}-E-i0^{+}}\right]^{-1},
\end{equation}
where $E\equiv\hbar^{2}k^{2}/M$. Using the relation $T^{-1}({\bf k},{\bf k};E)=-M[\cot\delta_{p}(k)-i]/(4\hbar^{2})$,
we find that in the limit $k\rightarrow0$,
\begin{equation}
\frac{1}{\lambda}+\mathcal{\mathscr{P}}\sum_{{\bf p}}\frac{\left|\Gamma\left({\bf p}\right)\right|^{2}}{2\epsilon_{{\bf p}}-E}=\frac{M\left|\Gamma\left({\bf k}\right)\right|^{2}}{4\hbar^{2}k^{2}}\left[\frac{1}{a_{p}}-\frac{2k^{2}}{\pi}\ln\left(R_{p}k\right)\right].
\end{equation}
As shown in Appendix A, for arbitrary exponent $n$ we have 
\begin{eqnarray}
\mathcal{\mathscr{P}}\sum_{{\bf p}}\frac{\left|\Gamma\left({\bf p}\right)\right|^{2}}{2\epsilon_{{\bf p}}-E} & = & \frac{M}{4\pi\hbar^{2}}\left[\frac{k_{0}^{2}}{k_{F}^{2}}\frac{\pi\left(n-1/2\right)\left(n-1\right)}{n^{3}\sin\left(\pi/n\right)}\right.\nonumber \\
 &  & \left.-\frac{2k^{2}}{k_{F}^{2}}\ln\left(e^{\frac{3}{4n}}\frac{k}{k_{0}}\right)\right].\label{eq:INT2BTMatrix}
\end{eqnarray}
Therefore, we obtain
\begin{eqnarray}
\frac{1}{a_{p}} & = & \frac{4\hbar^{2}k_{F}^{2}}{M\lambda}+\frac{\left(n-1/2\right)\left(n-1\right)}{n^{3}\sin\left(\pi/n\right)}k_{0}^{2},\\
R_{p} & = & \exp\left(\frac{3}{4n}\right)k_{0}^{-1}.
\end{eqnarray}
In the limit $n\rightarrow\infty$, we recover the known relation
$R_{p}=k_{0}^{-1}$ \cite{Hu2018}. For the expression of the scattering
area $a_{p}$, we may replace $\lambda$ in favor of the scattering
energy $E_{b}$. In the low-energy limit, i.e., $\left|E_{b}\right|\ll\hbar^{2}k_{0}^{2}/M$,
we find that, 
\begin{equation}
\frac{1}{a_{p}}=\frac{ME_{b}}{\pi\hbar^{2}}\left[\ln\frac{M\left|E_{b}\right|}{\hbar^{2}k_{0}^{2}}+\frac{3}{2n}\right].\label{eq:ap}
\end{equation}
It is easy to see that the scattering energy $E_{b}$ changes sign
in the unitary limit $a_{p}=\pm\infty$. On the BEC side with $a_{p}>0$,
we may write $E_{b}=-\varepsilon_{B}$, where $\varepsilon_{B}\equiv\hbar^{2}\kappa^{2}/M$
is the binding energy, and obtain
\begin{equation}
-\frac{1}{a_{p}}-\frac{2\kappa^{2}}{\pi}\ln\left(R_{p}\kappa\right)=0.\label{eq:kappa}
\end{equation}
This equation agrees with the low-energy expansion of the phase shift
in Eq. (\ref{eq:PhaseShift}), where $k=i\kappa$ is simply the pole
of the $p$-wave scattering amplitude $f_{p}(k)=\sqrt{2/(\pi k)}[\cot\delta_{p}(k)-i]^{-1}$.

\section{Zero temperature theory}

The zero-temperature mean-field and GPF theories of a 2D chiral $p$-wave
Fermi superfluid were laid out in our previous work \cite{Hu2018}.
Here, for self-containedness, we give a brief summary. In the superfluid
phase, two fermions can pair up via the separable attraction $V_{\mathbf{k}\mathbf{k}'}$
to form a Copper pair, described by a generalized density operator
$\hat{\rho}_{\mathbf{q}}\equiv\lambda\sum_{{\bf k}}\Gamma^{*}(\mathbf{k})b_{\mathbf{kq}}$.
The pairs then condense into the zero center-of-mass momentum state,
as described by a nonzero pairing order parameter $\Delta$, i.e.,

\begin{equation}
\hat{\rho}_{\mathbf{q}}=\Delta\delta_{\mathbf{q},\mathbf{0}}+\Delta_{\mathbf{q}}.
\end{equation}
On top of this condensate are strong pair fluctuations, represented
by the field operator $\Delta_{\mathbf{q}}$ for the non-condensed
Cooper pairs.

Neglecting $\Delta_{\mathbf{q}}$ leads to the mean-field description.
At a given chemical potential, the zero-temperature thermodynamic
potential takes the form \cite{Hu2018},
\begin{equation}
\Omega_{\text{MF}}=\frac{1}{2}\frac{\Delta^{2}}{\lambda}+\frac{1}{2}\sum_{{\bf k}}\left(\xi_{{\bf k}}-E_{{\bf k}}\right),
\end{equation}
where $E_{{\bf k}}=[\xi_{{\bf k}}^{2}+\Delta^{2}\left|\Gamma({\bf k})\right|^{2}]^{1/2}$
is the energy of fermionic Bogoliubov quasi-particles. The associated
quasi-particle wave functions are given by,
\begin{eqnarray}
\left|u_{{\bf k}}\right|^{2} & = & \frac{1}{2}\left(1+\frac{\xi_{{\bf k}}}{E_{{\bf k}}}\right),\\
\left|v_{{\bf k}}\right|^{2} & = & \frac{1}{2}\left(1-\frac{\xi_{{\bf k}}}{E_{{\bf k}}}\right),\\
u_{{\bf k}}v_{{\bf k}}^{*} & = & \frac{\Delta\Gamma\left({\bf k}\right)}{2E_{{\bf k}}}.
\end{eqnarray}
By minimizing the mean-field thermodynamic potential with respect
to $\Delta$ and $\mu$, we obtain the mean-field gap equation,
\begin{equation}
\frac{1}{\lambda}+\sum_{{\bf k}}\frac{\left|\Gamma\left({\bf k}\right)\right|^{2}}{2E_{{\bf k}}}=0,\label{eq:gapMF}
\end{equation}
and the mean-field number equation,
\begin{equation}
n_{2D}=-\frac{\partial\Omega_{\text{MF}}}{\partial\mu}=\frac{1}{2}\sum_{{\bf k}}\left(1-\frac{\xi_{{\bf k}}}{E_{{\bf k}}}\right)\equiv n_{F}.\label{eq:numMF}
\end{equation}

The contribution of strong pair fluctuations to the thermodynamic
potential can be accounted for, by taking an approximate Green function
$\Gamma(\mathcal{Q}\equiv\left\{ \mathbf{q},i\nu_{n}\right\} )$ for
non-condensed Copper pairs at the Gaussian level \cite{Hu2006},
\begin{equation}
\Gamma\left(\mathcal{Q}\right)=-\left[\begin{array}{ll}
M_{11}\left(\mathcal{Q}\right) & M_{12}\left(\mathcal{Q}\right)\\
M_{21}\left(\mathcal{Q}\right) & M_{22}\left(\mathcal{Q}\right)
\end{array}\right]^{-1},
\end{equation}
where the matrix elements are given by,\begin{widetext}
\begin{eqnarray}
M_{11}\left(\mathcal{Q}\right) & = & \sum_{{\bf k}}\left|\Gamma_{\mathbf{}}\left({\bf k}\right)\right|^{2}\left[\frac{\left(u_{+}u_{+}^{*}\right)\left(u_{-}u_{-}^{*}\right)}{i\nu_{n}-E_{+}-E_{-}}-\frac{\left(v_{+}v_{+}^{*}\right)\left(v_{-}v_{-}^{*}\right)}{i\nu_{n}+E_{+}+E_{-}}+\frac{1}{2E_{\mathbf{k}}}\right],\\
M_{12}\left(\mathcal{Q}\right) & = & \sum_{{\bf k}}\left[\Gamma^{*}\left({\bf k}\right)\right]^{2}\left[\frac{\left(u_{+}v_{+}^{*}\right)\left(u_{-}v_{-}^{*}\right)}{i\nu_{n}-E_{+}-E_{-}}-\frac{\left(u_{+}v_{+}^{*}\right)\left(u_{-}v_{-}^{*}\right)}{i\nu_{n}+E_{+}+E_{-}}\right],
\end{eqnarray}
\end{widetext}$M_{21}(\mathcal{Q})=M_{12}^{*}(\mathcal{Q})$, and
$M_{22}(\mathcal{Q})=M_{11}^{*}(\mathcal{Q})$. Here, $\nu_{n}\equiv2n\pi k_{B}T$
with integer $n=0,\pm1,\pm2,\cdots$ are bosonic Matsubara frequencies,
and the abbreviations $u_{\pm}\equiv u_{{\bf q}/2\pm{\bf k}}$, $v_{\pm}\equiv v_{{\bf q}/2\pm{\bf k}}$,
and $E_{\pm}\equiv E_{{\bf q}/2\pm{\bf k}}$ are used. At the Gaussian
level, the effective interaction between non-condensed Cooper pairs
is treated within the Bogoliubov approximation, so there is no residual
interaction between bosonic quasi-particles. Therefore, it is straightforward
to write down the fluctuation part of the thermodynamic potential
for non-interacting quasi-particles \cite{Abrikosov1963},

\begin{equation}
\Omega_{\text{GF}}\left[\mu,\Delta\left(\mu\right)\right]=\frac{k_{B}T}{2}\sum_{i\nu_{n}}\sum_{\mathbf{q}}\ln\det\left[-\Gamma^{-1}\left(\mathcal{Q}\right)\right].\label{eq:OmegaGF}
\end{equation}
For a given $\mu$, once $\Omega_{\textrm{GF}}$ is numerically calculated,
we determine the number of Cooper pairs $n_{B}$ by using numerical
differentiation, 
\begin{equation}
2n_{B}=-\frac{\partial\Omega_{\textrm{GF}}\left[\mu,\Delta\left(\mu\right)\right]}{\partial\mu}.
\end{equation}
The number equation Eq. (\ref{eq:numMF}) is then updated to, 
\begin{equation}
n_{2D}=n_{F}+2n_{B}.
\end{equation}
This leads to an updated chemical potential in the GPF theory. 

It is worth noting that within GPF the pairing gap $\Delta\left(\mu\right)$
is always calculated at the mean-field level, by solving the gap equation
Eq. (\ref{eq:gapMF}). This is necessary to ensure a gapless Goldstone
phonon mode \cite{Hu2006,Diener2008}, i.e., $\det\Gamma^{-1}(\mathcal{Q}=0)=0$.
In principle, it is possible to have a generalized approximation to
improve the gap equation beyond mean-field. Accordingly, we could
improve the vertex function $\Gamma(\mathcal{Q})$ beyond GPF. This
possibility will be explored in future studies. 

\subsection{Analytic solutions from mean-field theory}

In two dimensions, the integrals involved in mean-field equations
can often be integrated out explicitly, leading to some nice analytic
solutions. In the following, we take Fermi wave-vector $k_{F}$ and
Fermi energy $\varepsilon_{F}=\hbar^{2}k_{F}^{2}/(2M)$ as the units
of wave-vector and energy, respectively. In particular, we define
the dimensionless pairing gap $\tilde{\Delta}=\Delta/\varepsilon_{F}$
and the dimensionless chemical potential $\tilde{\mu}=\mu/\varepsilon_{F}$.
By setting $n\rightarrow\infty$ (i.e., taking a step-like function
for the regularization function $\Gamma({\bf k})$) and performing
the integrals in the gap and number equations, we arrive at two coupled
equations:
\begin{equation}
-\left[\frac{\tilde{\Delta}^{2}}{4}-\tilde{\mu}\Theta\left(\tilde{\mu}\right)\right]+\frac{\tilde{\Delta}^{2}}{2}\ln\frac{(k_{F}R_{p})^{-1}}{\sqrt{\frac{\tilde{\Delta}^{2}}{4}-\tilde{\mu}\Theta\left(-\tilde{\mu}\right)}}=1\label{eq:MF1}
\end{equation}
and
\begin{equation}
\tilde{\mu}\ln\frac{(k_{F}R_{p})^{-1}}{\sqrt{\frac{\tilde{\Delta}^{2}}{4}-\tilde{\mu}\Theta\left(-\tilde{\mu}\right)}}=1-\frac{\pi}{2}\frac{1}{k_{F}^{2}a_{p}},\label{eq:MF2}
\end{equation}
where $\Theta(x)$ is the step function.

In the BCS limit, where $\tilde{\Delta}\rightarrow0$ and $\tilde{\mu}\rightarrow1$,
we find from Eq. (\ref{eq:MF2}) that 
\begin{equation}
\frac{\Delta}{\varepsilon_{F}}\simeq\frac{2}{e}\left(\frac{1}{k_{F}R_{p}}\right)\exp\left[\frac{\pi}{2}\frac{1}{k_{F}^{2}a_{p}}\right].\label{eq:GapBCS}
\end{equation}
By substituting it into Eq. (\ref{eq:MF1}), we obtain
\begin{equation}
\frac{\mu}{\varepsilon_{F}}\simeq1+\frac{\pi}{e^{2}}\left(\frac{1}{k_{F}R_{p}}\right)^{2}\frac{1}{k_{F}^{2}a_{p}}\exp\left[\frac{\pi}{k_{F}^{2}a_{p}}\right].
\end{equation}

\begin{figure}
\centering{}\includegraphics[width=0.48\textwidth]{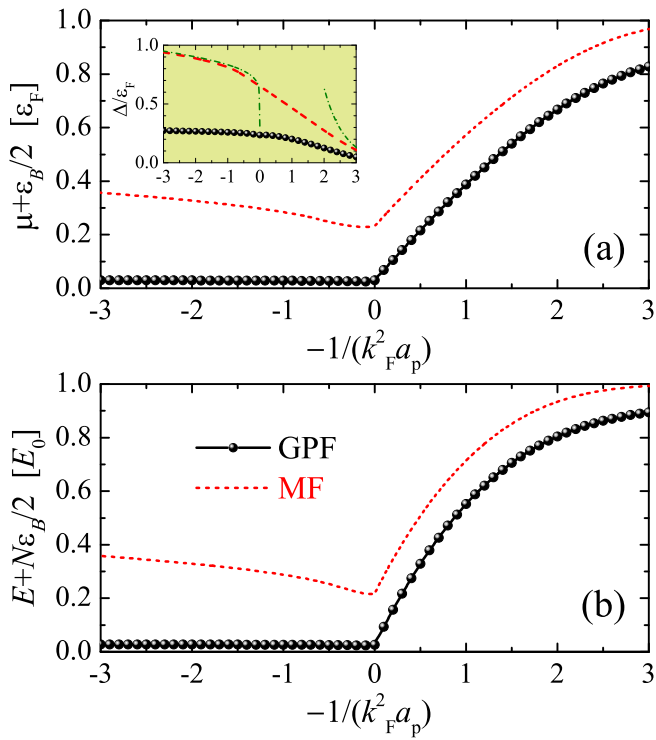}\caption{\label{fig1_EoS} The chemical potential $\mu$ (in units of $\varepsilon_{F}$)
and the total energy $E$ (in units of $E_{0}=N\varepsilon_{F}/2$),
as a function of the inverse scattering area $-1/(k_{F}^{2}a_{p})$,
calculated by using the mean-field theory (red dashed line) and the
GPF theory (black solid line with circles). We have subtracted the
contribution from the two-body bound state with binding energy $\varepsilon_{B}\equiv-E_{b}$
when the scattering area is positive. The effective range of the interaction
is fixed to $k_{F}R_{p}=0.05$. The inset in (a) shows the pairing
order parameter. The two green dot-dashed lines are the mean-field
predictions of the asymptotic behavior in the BCS and BEC limits,
Eqs. (\ref{eq:GapBCS}) and (\ref{eq:GapBEC}), respectively. }
\end{figure}

In the BEC limit, the chemical potential becomes negative and approaches
the half of the bound state energy, i.e., $\mu\rightarrow-\hbar^{2}\kappa^{2}/(2M)$,
where $\kappa$ is the solution of Eq. (\ref{eq:kappa}). From Eq.
(\ref{eq:MF1}), it is readily seen that,
\begin{equation}
\frac{\Delta}{\varepsilon_{F}}\simeq\left[-\frac{1}{2}\ln\left(R_{p}\kappa\right)-\frac{1}{4}\right]^{-1/2}.\label{eq:GapBEC}
\end{equation}
For the chemical potential, we rewrite it into the form, $\mu=-\hbar^{2}\kappa^{2}/(2M)+\mu_{B}/2$,
where the molecular chemical potential $\mu_{B}$ is approximately
equal to $g_{B}n_{2D}/2$ with $g_{B}$ being the strength of the
interaction between two pairs. After some algebra, we find that
\begin{equation}
\frac{\mu_{B}}{\varepsilon_{F}}\simeq\left[-\frac{1}{2}\ln\left(R_{p}\kappa\right)-\frac{1}{4}\right]^{-1}\simeq\left(\frac{\Delta}{\varepsilon_{F}}\right)^{2}.
\end{equation}
The pair-pair interaction strength $g_{B}=\mu_{B}/(n_{2D}/2)$ is
then given by,
\begin{equation}
g_{B}\simeq\frac{8\pi\hbar^{2}/M}{-\ln\left(R_{p}\kappa\right)}=\frac{16\pi\hbar^{2}/M}{\ln\left[\hbar^{2}k_{0}^{2}/(M\left|E_{b}\right|)\right]},\label{eq:gBMF}
\end{equation}
in agreement with the previous result (see Eq. (48) in Ref. \cite{Hu2018}).

\subsection{Numerical results on equation of state}

At the level beyond mean-field, the GPF theory can only be solved
numerically. In Fig. \ref{fig2_kFRp005contact}, we report the GPF
chemical potential (a), total energy (b) and the pairing gap (i.e.,
the inset) as a function of the inverse scattering area $-1/(k_{F}^{2}a_{p})$
at a given effective range $k_{F}R_{p}=0.05$, using the black solid
lines with circles. For comparison, we show also the corresponding
mean-field results by the red dashed lines. These thermodynamic variables
have been shown in the previous work as a function of the scattering
energy $E_{b}$ \cite{Hu2018}.

Both chemical potential and total energy suppress significantly from
their non-interacting values $\varepsilon_{F}$ and $E_{0}=N\varepsilon_{F}/2$,
respectively. In particular, on the BEC side, we observe a flat molecular
chemical potential and total energy, which are nearly independent
on the scattering area $a_{p}$. As discussed in the previous work
\cite{Hu2018}, this is an indication of the formation of an interacting
Bose condensate of composite Copper pairs in two dimensions, with
a constant pair-pair interaction strength $g_{B}\sim\hbar^{2}/M$.

\subsection{Super-Efimov trimers}

It is worth mentioning that, for 2D fermions with $p$-wave interaction,
Nishida and co-workers recently discovered a series of three-particle
bound states, namely super-Efimov states \cite{Nishida2013}. How
would the many-body properties of the system (i.e., contact and breathing
mode as addressed in this work) be affected by these super-Efimov
trimers is an interesting research topic to explore \cite{Zhang2017SuperEfimov}.
Naïvely, due to the double exponential scaling of the super-Efimov
trimers \cite{Nishida2013}, we anticipate that only one trimer with
an emergent energy scale will appear under current experimental conditions.
The neighboring trimer with smaller energy cannot exist due to its
large spatial extent, while the one with larger energy is simply too
deep to be experimentally observed. In this respect, the impact of
super-Efimov states to the many-body physics could be less significant
than that of conventional Efimov states.

\section{Results and discussions}

We are now ready to discuss the $p$-wave contacts and the related
breathing mode frequency. There are two contact parameters, characterizing
the short-distance and large-momentum behaviors of different correlation
functions, such as momentum distribution and pair-pair correlation
function \cite{Yoshida2015,Yu2015}. As shown by Yi-Cai Zhang and
Shizhong Zhang \cite{Zhang2017}, these two contacts $C_{a}$ and
$C_{R}$ satisfy the adiabatic relations,
\begin{eqnarray}
\left(\frac{\partial E}{\partial a_{p}^{-1}}\right)_{S} & = & -\frac{\pi\hbar^{2}}{2M}C_{a},\\
\left(\frac{\partial E}{\partial\ln R_{p}}\right)_{S} & = & \frac{\hbar^{2}}{M}C_{R},\label{eq:AdiabaticRelationCR}
\end{eqnarray}
 and therefore can be determined once the total energy is known at
a given entropy $S$. At zero temperature, where the entropy is always
zero, we simply take the two first-order derivatives.

\subsection{Tan's $p$-wave contacts}

For this purpose, we may write the zero-temperature total energy in
a dimensionless form $\xi(x,y)$,
\begin{equation}
E=\frac{N\varepsilon_{F}}{2}\xi\left[x\equiv\frac{1}{k_{F}^{2}a_{p}},y\equiv\ln\left(k_{F}R_{p}\right)\right],
\end{equation}
and the two $p$-wave contacts can similarly be rewritten in the dimensionless
way,
\begin{eqnarray}
\frac{C_{a}}{N} & = & -\frac{1}{2\pi}\xi_{x},\label{eq:Ca}\\
\frac{C_{R}}{Nk_{F}^{2}} & = & \frac{1}{4}\xi_{y},\label{eq:CR}
\end{eqnarray}
where $\xi_{x}\equiv\partial\xi/\partial x$ and $\xi_{y}\equiv\partial\xi/\partial y$.
Following the dimensionless form of the total energy, it is easy to
find that the chemical potential $\mu=\partial E/\partial N$ and
the pressure $P=\mu n_{2D}-E/A$,
\begin{eqnarray}
\mu & = & \varepsilon_{F}\left(\xi-\frac{x}{2}\xi_{x}+\frac{1}{4}\xi_{y}\right),\\
P & = & P_{0}\left(\xi-x\xi_{x}+\frac{1}{2}\xi_{y}\right),\label{eq:PressureEOS}
\end{eqnarray}
where $P_{0}\equiv n_{2D}\varepsilon_{F}/2$. By substituting the
expressions of the dimensionless contact into the last equation for
pressure, we obtain the pressure relation \cite{Zhang2017},
\begin{equation}
PA=E+\frac{\pi\hbar^{2}}{2M}\frac{C_{a}}{a_{p}}+\frac{\hbar^{2}}{2M}C_{R}.
\end{equation}

It is useful to distinguish the two- and many-body contributions to
the contact parameters. For the two-body contribution, we assume that
the system can be viewed as an ideal, non-interacting gas of $N/2$
pairs, each of which has the energy,
\begin{equation}
\varepsilon_{2B}=\left\{ \begin{array}{cc}
-\varepsilon_{B}, & \textrm{if \ensuremath{a_{p}>0}}\\
0, & \textrm{otherwise}
\end{array}\right..
\end{equation}
 In other words, on the BEC side the pair takes the ground-state energy
of the two-body bound state; while on the BCS side, the minimum energy
of the pair should be zero (i.e., the lower threshold of the two-particle
continuum). The two-body contribution to the total energy of the system
can then be written as, 
\begin{equation}
E_{2B}=\left(N/2\right)\varepsilon_{2B}=\left\{ \begin{array}{cc}
-N\varepsilon_{B}/2, & \textrm{if \ensuremath{a_{p}>0}}\\
0, & \textrm{otherwise}
\end{array}\right..
\end{equation}
On the BEC side, by using Eq. (\ref{eq:kappa}), we find that the
two-body contribution to the contact from the energy $E_{2B}$, denoted
by $C_{a,2B}$ and $C_{R,2B}$, is given by \cite{Zhang2017},
\begin{eqnarray}
C_{a,2B} & = & -\frac{N}{2}\frac{1}{\ln\left(R_{p}\kappa\right)+1/2},\\
C_{R,2B} & = & +\frac{N}{2}\frac{\kappa^{2}}{\ln\left(R_{p}\kappa\right)+1/2}.
\end{eqnarray}
On the BCS side, $C_{a,2B}=0$ and $C_{R,2B}=0$, as a result of $E_{2B}=0$.
On both sides, either BEC or BCS, we obtain that,
\begin{equation}
E_{2B}+\frac{\pi\hbar^{2}}{2M}\frac{C_{a,2B}}{a_{p}}+\frac{\hbar^{2}}{2M}C_{R,2B}=0.
\end{equation}
This equation is easy to understand from the pressure relation, since
the two-body bound state does not contribute to the many-body observables
such as pressure.

\begin{figure}
\centering{}\includegraphics[width=0.48\textwidth]{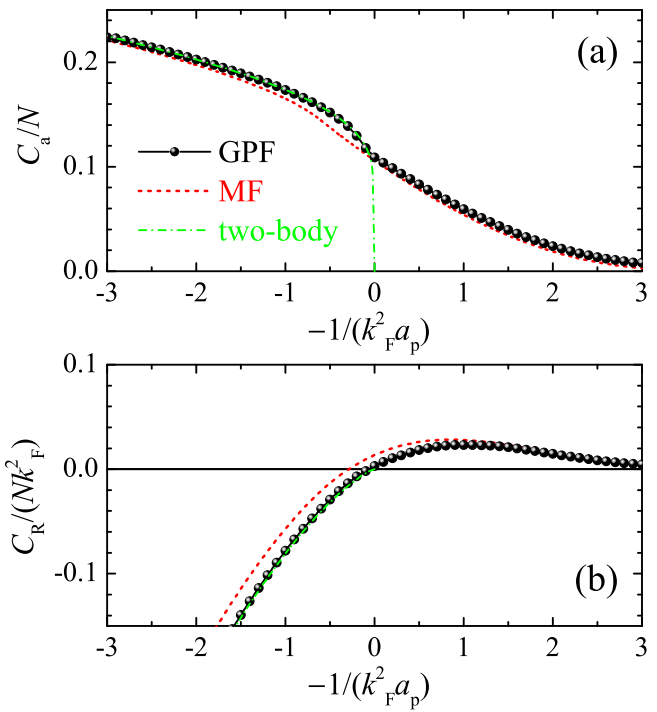}\caption{\label{fig2_kFRp005contact} (color online). The contact parameters
$C_{a}$ (a) and $C_{R}$ (b) as a function of the inverse scattering
area $-1/(k_{F}^{2}a_{p})$ at the effective range of the interaction
$k_{F}R_{p}=0.05$. The mean-field and GPF results are shown by the
red dashed line and the black solid line with circles, respectively.
The two-body contribution is also shown by the green dot-dashed line. }
\end{figure}

In Fig. \ref{fig2_kFRp005contact}, we plot the two dimensionless
contact parameters as a function of $-1/(k_{F}^{2}a_{p})$ at a given
effective range $k_{F}R_{p}=0.05$, calculated by using the GPF theory
(black lines with circles) and the mean-field theory (red dashed lines).
The two-body contribution from the bound state to the contacts is
also shown by green dot-dashed lines. As we see in Fig. \ref{fig2_kFRp005contact}(a),
the contact related to the scattering area $C_{a}$ is always positive.
It increases with increasing interaction strength $(k_{F}^{2}a_{p})^{-1}$.
On the BEC side with a positive scattering area, we find that the
GPF result of $C_{a}$ is exhausted by the two-body contribution $C_{a,2B}$.
The mean-field theory seems to under-estimate $C_{a}$, with the largest
under-estimation occurs at $(k_{F}^{2}a_{p})^{-1}\sim0.5$. On the
other hand, the contact related to the effective range, $C_{R}$,
has a non-monotonic dependence on the inverse scattering area, as
shown in Fig. \ref{fig2_kFRp005contact}(b). As $(k_{F}^{2}a_{p})^{-1}$
increases, $C_{R}$ initially increases, reaches a maximum at $(k_{F}^{2}a_{p})^{-1}\sim-1$,
and then decreases to zero at about the resonance limit. Towards the
BEC limit, it decreases very rapidly. We find similarly that the GPF
result of $C_{R}$ is almost exhausted by the two-body contribution
$C_{R,2B}$. The mean-field theory generally over-estimates $C_{R}$
and the over-estimation becomes increasingly larger when we increase
interaction strength. This is related to the unreliable prediction
of the mean-field theory on the pair-pair interaction strength (see
Eq. (\ref{eq:gBMF})).

\begin{figure}
\centering{}\includegraphics[width=0.48\textwidth]{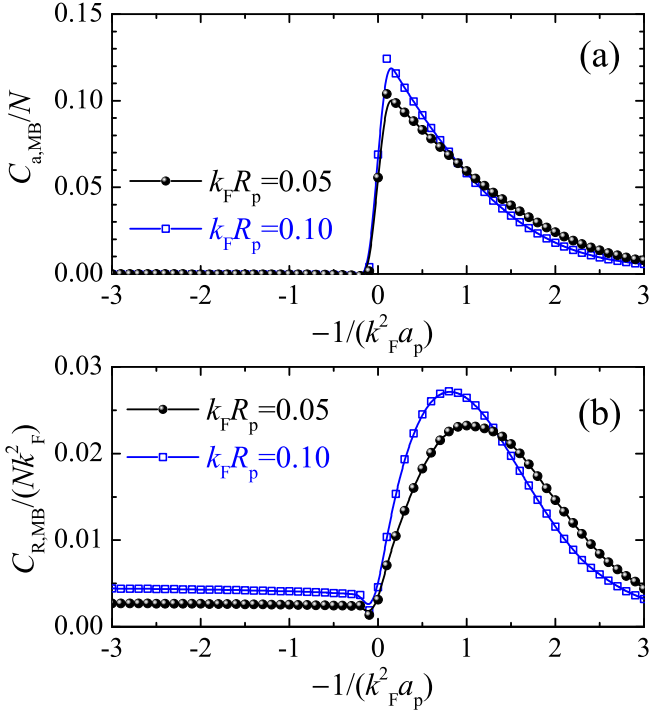}\caption{\label{fig3_contactMB} (color online). The many-body part of the
contact parameters, $C_{a,MB}$ (a) and $C_{R,MB}$ (b), as a function
of the inverse scattering area $-1/(k_{F}^{2}a_{p})$ at two effective
ranges of the interaction $k_{F}R_{p}=0.05$ (black solid line with
circles) and $k_{F}R_{p}=0.10$ (blue solid line with squares). All
the results are calculated by using the GPF theory.}
\end{figure}

We have separated out the many-body parts of the two contact parameters,
$C_{a,MB}=C_{a}-C_{a,2B}$ and $C_{R,MB}=C_{R}-C_{R,2B}$, and show
the GPF predictions in Fig. \ref{fig3_contactMB}, for two effective
ranges of interactions, $k_{F}R_{p}=0.05$ (black line with circles)
and $k_{F}R_{p}=0.10$ (blue line with squares). On the BEC side,
the many-body parts of both contact parameters are small, consistent
with the observation in Fig. \ref{fig2_kFRp005contact} that the contacts
are exhausted by the two-body contribution. Across the resonance limit,
they exhibit a pronounced peak. The peak in $C_{a,MB}$ is slightly
above the resonance limit. The peak in $C_{R,MB}$ locates at $(k_{F}^{2}a_{p})^{-1}\sim-1$
and shifts towards the BCS limit with decreasing effective range.
We note that the many-body parts of the two $p$-wave contacts are
always positive.

\subsection{Breathing mode frequency}

The interesting dependence of the many-body part of the contacts on
the interaction strength may lead to a non-trivial breathing-type
oscillation mode, when the 2D $p$-wave Fermi superfluid is confined
in a harmonic trap with trapping frequency $\omega_{0}$. This is
a mode excited by the perturbation $\lambda(t)\mathcal{O}\equiv\lambda(t)\sum_{i=1}^{N}r_{i}^{2}$,
i.e., by slightly modulating the harmonic trapping frequency for a
certain period. For non-interacting bosons or fermions, the breathing
mode frequency is simply $\omega_{c}=2\omega_{0}$. The inter-particle
interaction generally leads to a frequency shift. As shown by Yi-Cai
Zhang and Shizhong Zhang \cite{Zhang2017}, the frequency shift at
the leading order is proportional to the contact parameters. By using
virial expansion, the frequency shift of the breathing mode at high
temperatures was then theoretically studied \cite{Zhang2017}.

In our zero-temperature case, we calculate the breathing mode frequency
using the well-known scaling approach \cite{Menotti2002,Hu2004,Hu2014}.
This amounts to assuming a polytropic form for the pressure equation
of state, $P\propto n_{2D}^{\gamma+1}$, where the polytropic index
$\gamma$ may be calculated using
\begin{equation}
\gamma=\frac{n_{2D}}{P}\left(\frac{\partial P}{\partial n_{2D}}\right)-1,\label{eq:PolyGamma}
\end{equation}
at the center of the harmonic trap. The scaling approach then leads
to a breathing mode frequency \cite{Hu2014},
\begin{equation}
\frac{\omega_{B}^{2}}{\omega_{c}^{2}}=\frac{\gamma+1}{2}=\frac{P_{0}\kappa_{T}^{(0)}}{P\kappa_{T}},\label{eq:BMFreq}
\end{equation}
where at zero temperature we rewrite $\partial P/\partial n_{2D}$
in terms of the compressibility $\kappa_{T}=[n_{2D}^{2}(\partial\mu/\partial n_{2D})]^{-1}$
and $\kappa_{T}^{(0)}\equiv(n_{2D}\varepsilon_{F})^{-1}$ is its non-interacting
value. This expression emphasizes the sound-wave nature of the breathing
mode frequency. Qualitatively, the breathing mode frequency can be
estimated as $c_{s}k_{\textrm{min}}$, where $c_{s}$ is the sound
velocity and $k_{\textrm{min}}\propto A^{-1/2}$ is the minimum wavevector
of the Fermi cloud with an area $A$. By recalling the relation $\kappa_{T}^{-1}\propto c_{s}^{2}$
and assuming the pressure $P\propto A^{-1}$ under the soft-wall confinement
of the harmonic trap, we find $\omega_{B}^{2}\propto(P\kappa_{T})^{-1}$.

The polytropic index $\gamma$ can be directly calculated once the
energy or pressure equation of state is known. By taking derivative
with respect to density in Eq. (\ref{eq:PressureEOS}) and neglecting
all small second-order derivatives, we find that,
\begin{equation}
n_{2D}\frac{\partial P}{\partial n_{2D}}\simeq2P+P_{0}\left(-x\bar{\xi}_{x}+\frac{1}{2}\bar{\xi}_{y}\right),
\end{equation}
where the \emph{bar} over $\xi_{x}$ and $\xi_{y}$ indicates that
we do not include the irrelevant two-body contribution. By substituting
it into Eq. (\ref{eq:PolyGamma}), we obtain the frequency shift $\delta\omega_{B}=\omega_{B}-\omega_{c}$,
\begin{equation}
\frac{\delta\omega_{B}}{\omega_{c}}\simeq\frac{\gamma-1}{4}\simeq\frac{-x\bar{\xi}_{x}+\frac{1}{2}\bar{\xi}_{y}}{4P/P_{0}}.
\end{equation}
By replacing the two derivatives with the help of Eqs. (\ref{eq:Ca})
and (\ref{eq:CR}), we finally arrive at
\begin{equation}
\frac{\delta\omega_{B}}{\omega_{c}}\simeq\frac{\hbar^{2}}{M^{2}}\frac{\left[\pi a_{p}^{-1}C_{a,MB}+C_{R,MB}\right]}{4\omega_{0}^{2}\left\langle \mathcal{O}\right\rangle },\label{eq:FreqShiftContact}
\end{equation}
where the virial theorem $PA=M\omega_{0}^{2}\left\langle \mathcal{O}\right\rangle /2$
in the presence of harmonic traps is used \cite{NoteVirialTheorem}.
We therefore recover Eq. (84) in Ref. \cite{Zhang2017} and explicitly
show the relation between the many-body parts of the two $p$-wave
contacts and the frequency shift in the breathing mode.

\begin{figure}
\begin{centering}
\includegraphics[width=0.48\textwidth]{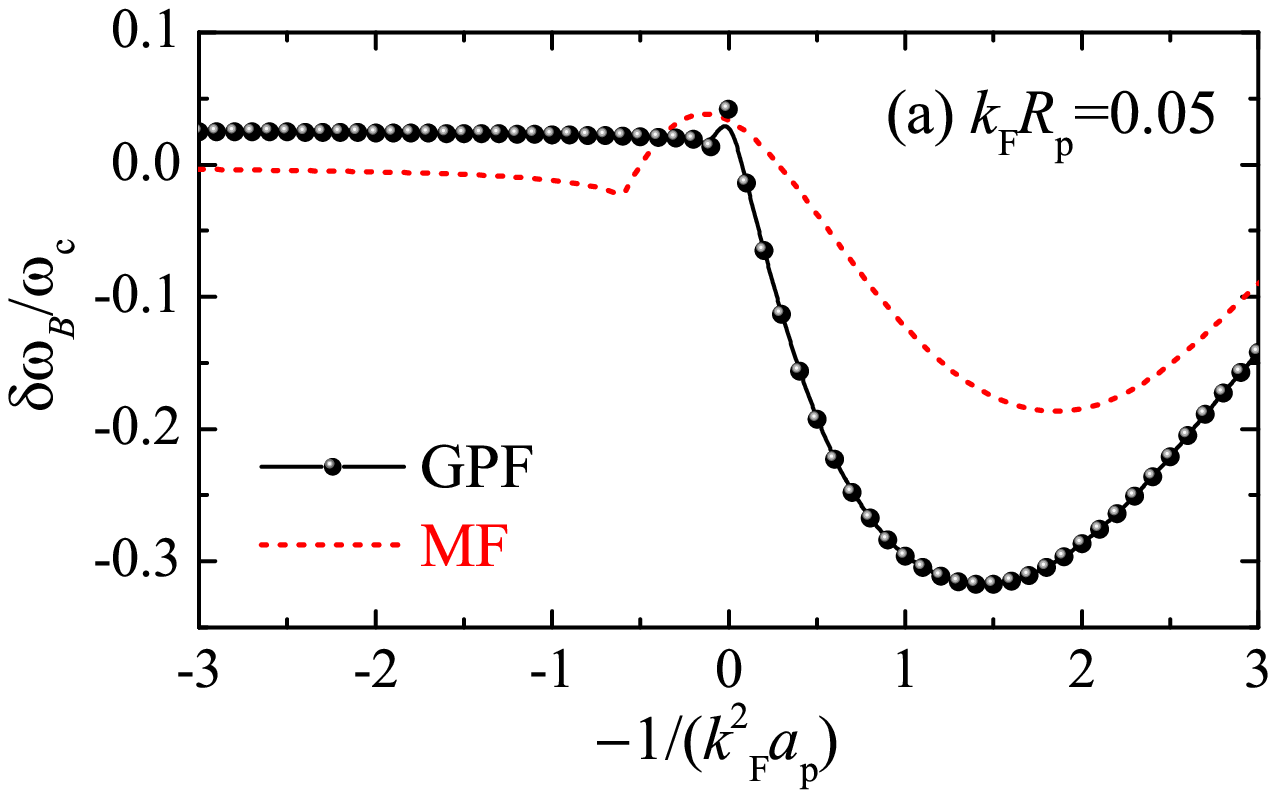}
\par\end{centering}
\centering{}\includegraphics[width=0.48\textwidth]{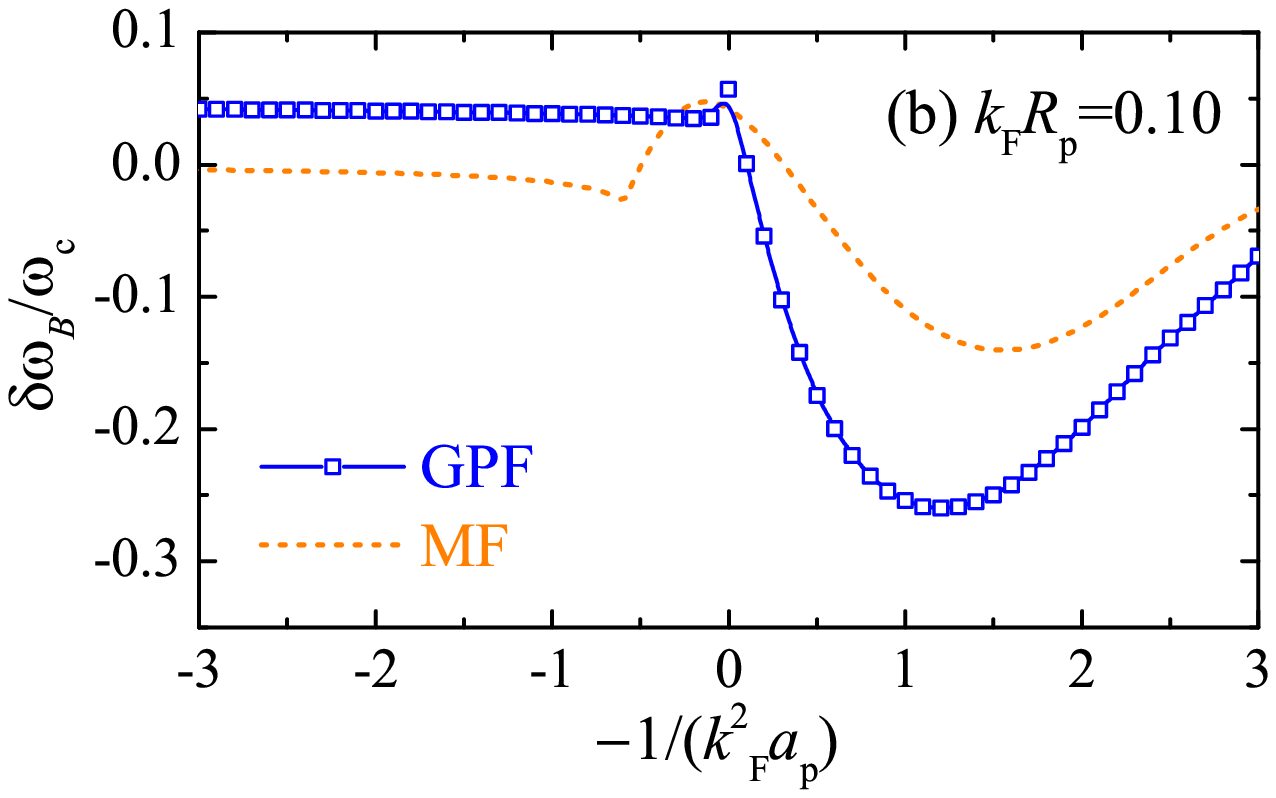}\caption{\label{fig4_wb} (color online). The deviation of the breathing mode
frequency from the scale-invariant result of $\omega_{c}=2\omega_{0}$,
as a function of the inverse scattering area $-1/(k_{F}^{2}a_{p})$,
at two effective ranges of the interaction: $k_{F}R_{p}=0.05$ (a)
and $k_{F}R_{p}=0.10$ (b). The mean-field and GPF results are shown
by the dashed lines and the solid lines with symbols, respectively.}
\end{figure}

In Fig. \ref{fig4_wb}, we present the frequency shift of the breathing
mode as a function of the inverse scattering area, at two effective
ranges $k_{F}R_{p}=0.05$ (a) and $k_{F}R_{p}=0.10$ (b). These results
are calculated using Eq. (\ref{eq:BMFreq}) within the mean-field
theory (dashed lines) and the GPF theory (lines with symbols). We
find that the frequency shift is negative on the BCS side and exhibits
a broad dip at $-(k_{F}^{2}a_{p})^{-1}\sim1-2$. This dip structure
is apparently related to the peak structure in the many-body part
of the two $p$-wave contacts, according to Eq. (\ref{eq:FreqShiftContact}).
The two contacts contribute differently in opposite signs and the
contribution from $C_{a,MB}$ seems to dominate. We note that, in
a 1D harmonically trapped $p$-wave Fermi superfluid, the breathing
mode frequency shows qualitatively similar dependence on the interacting
strength in the weak-coupling regime \cite{Imambekov2010,Chen2016}.

On the BEC side, we see that the frequency shift predicted by the
GPF theory becomes flat and small. This is associated with the formation
of tight-binding molecules who interact via a nearly constant molecular
scattering length, as we discussed earlier. The GPF frequency shift
is positive and is about 5\% at $k_{F}R_{p}=0.10$. In contrast, the
mean-field frequency shift is negative and shows a non-trivial cusp
at $(k_{F}^{2}a_{p})^{-1}\sim0.5$. This mean-field behavior is unphysical,
arising from the unreliable equations of state predicted by the mean-field
theory. It is interesting to note that, the breathing mode frequency
shift of a weakly-interacting 2D Bose gas was investigated both theoretically
and experimentally \cite{Olshanii2010,Merloti2013}. In that case,
the shift is too small to be experimentally observed. The moderately
interacting molecular condensate formed in the strongly-interacting
2D $p$-wave Fermi superfluid could be a possible candidate to observe
the breathing mode frequency shift due to beyond-mean-field effects.

\section{Frequency shift in the resonance limit }

Here we focus on the breathing mode frequency in the resonance limit
$a_{p}=\pm\infty$. If we neglect the dependence of the equations
of state on the effective range, the dimensionless energy function
$\xi$ is simply a constant. From the pressure $P=\xi P_{0}\propto n_{2D}^{2}$,
we find a polytropic index $\gamma=2$ and hence $\omega_{B}=2\omega_{0}$.
This could be an exact result ensured by the scale invariance of the
system \cite{Werner2006}. However, the \emph{necessary} existence
of the effective range breaks the scale invariance and leads to a
derivation of the breathing mode frequency away from the scale-invariant
result of $\omega_{c}=2\omega_{0}$. A similar situation happens in
an $s$-wave 2D Fermi superfluid \cite{Hofmann2012,Vogt2012}. While
the superfluid with $s$-wave contact interaction is scale invariant
in the classical treatment, i.e., the model Hamiltonian simply scales
upon stretching the length of the system \cite{Pitaevskii1997}, the
renormalization of the contact interaction necessarily introduces
a 2D $s$-wave scattering length $a_{2D}$ that violates the scale-invariance.
This leads to an up-shift in the breathing mode frequency, the so-called
quantum anomaly, which is about 10\% in the strongly-interacting regime
\cite{Hofmann2012,Taylor2012}. It is not a surprise to see the similarity
between the effective range $R_{p}$ in a $p$-wave Fermi superfluid
and the 2D scattering length $a_{2D}$ in an $s$-wave Fermi superfluid.
This is discussed in more detail in Appendix B.

\begin{figure}
\centering{}\includegraphics[width=0.48\textwidth]{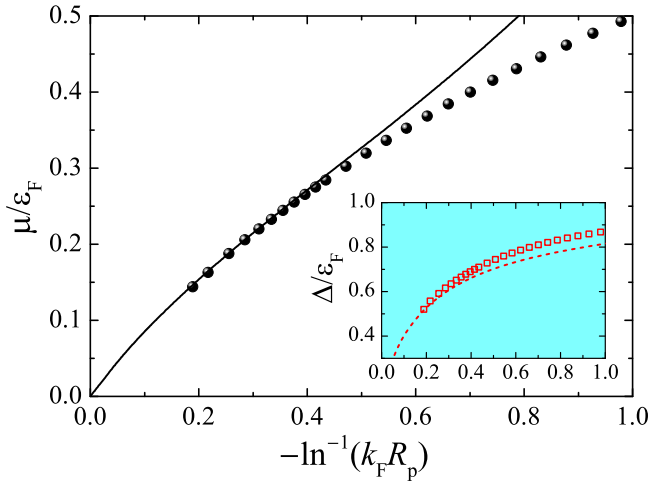}\caption{\label{fig5_unitaryEoSMF} (color online). The mean-field chemical
potential $\mu$ (main figure, circles) and the mean-field pairing
order parameter (inset, squares) as a function of $-\ln^{-1}(k_{F}R_{p})$
in the resonance limit (i.e., $a_{p}\rightarrow\pm\infty$). The lines
show the asymptotic behavior in the limit of zero effective range
of the interaction, $R_{p}\rightarrow0$ or $-\ln^{-1}(k_{F}R_{p})\rightarrow0$,
see, Eqs. (\ref{eq:UnitaryMuMF}) and (\ref{eq:UnitaryGapMF}).}
\end{figure}

\begin{figure}
\centering{}\includegraphics[width=0.48\textwidth]{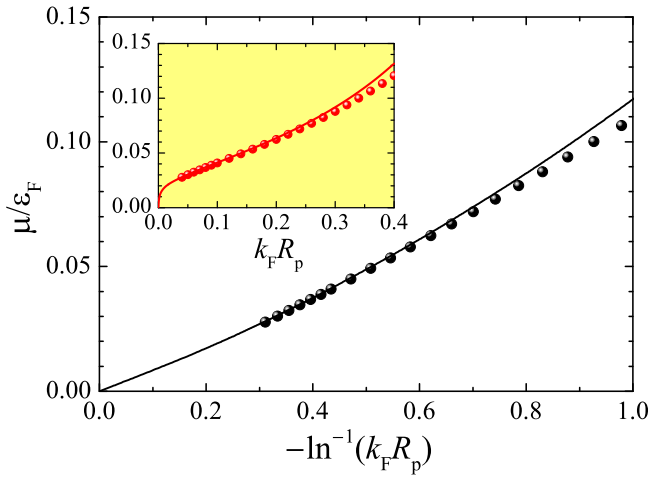}\caption{\label{fig6_unitaryEoSGPF} (color online). The chemical potential
$\mu$ as a function of $-\ln^{-1}(k_{F}R_{p})$, calculated by using
the GPF theory in the resonance limit $a_{p}\rightarrow\pm\infty$.
The inset shows the chemical potential as a function of $k_{F}R_{p}$
in the linear scale. The lines are the fitting curves to the GPF results,
see Eq. (\ref{eq:UnitaryMuGPF}).}
\end{figure}

\subsection{Chemical potential in the resonance limit}

Before we discuss the frequency shift in a resonantly interacting
$p$-wave Fermi superfluid, it is useful to first understand the chemical
potential in this limit. Using the mean-field equations Eqs. (\ref{eq:gapMF})
and (\ref{eq:numMF}), we find that,
\begin{eqnarray}
\tilde{\mu} & = & -\frac{1}{\ln\left(k_{F}R_{p}\tilde{\Delta}/2\right)},\\
\tilde{\Delta}^{2} & = & \frac{2\tilde{\mu}\left(1-\tilde{\mu}\right)}{1-\tilde{\mu}/2}.
\end{eqnarray}
By treating $-[\ln(k_{F}R_{p})]^{-1}=-y^{-1}$ as the small parameter,
we obtain,
\begin{eqnarray}
\frac{\mu}{\varepsilon_{F}} & \simeq & -\frac{1}{y}\left[1-\frac{1}{2y}\ln\left(-\frac{1}{2y}\right)\right],\label{eq:UnitaryMuMF}\\
\frac{\Delta}{\varepsilon_{F}} & \simeq & \sqrt{-\frac{2}{y}}\left[1-\frac{1}{4y}\ln\left(-\frac{1}{2e}\frac{1}{y}\right)\right].\label{eq:UnitaryGapMF}
\end{eqnarray}
Thus, towards the zero-range limit, the mean-field chemical potential
at resonance vanishes linearly.

More accurate predictions from the GPF theory should be determined
numerically. Empirically, we find that the GPF chemical potential
at resonance can be nicely fitted by the formalism,
\begin{equation}
\frac{\mu}{\varepsilon_{F}}\simeq-A\left(\frac{1}{y}-\frac{1}{2y^{2}}\right),\label{eq:UnitaryMuGPF}
\end{equation}
where $A\simeq0.078\ll1$. While the GPF chemical potential at resonance
still vanishes linearly in the zero-range limit, the slope (i.e.,
the value of $A$) is much slower than that of the mean-field chemical
potential.

In Figs. (\ref{fig5_unitaryEoSMF}) and (\ref{fig6_unitaryEoSGPF}),
we report the mean-field and GPF predictions of the chemical potential
at resonance as a function of $-\ln^{-1}(k_{F}R_{p})$, respectively.
The analytic expressions and the empirical formalism discussed in
the above are also shown. At small effective range, they agree well
with the numerical results. 

\begin{figure}
\centering{}\includegraphics[width=0.48\textwidth]{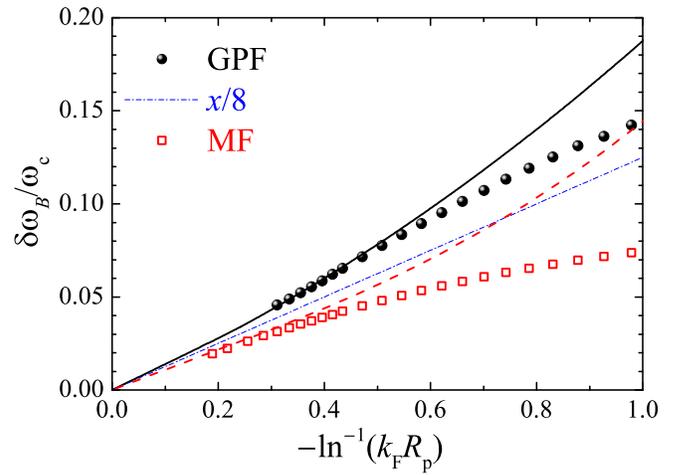}\caption{\label{fig7_unitarywb} (color online). The deviation of the breathing
mode frequency from the scale-invariant result of $\omega_{c}=2\omega_{0}$
as a function of $-\ln^{-1}(k_{F}R_{p})$ in the resonance limit $a_{p}\rightarrow\pm\infty$.
The mean-field and GPF results are shown by the black circles and
red squares, respectively. The black solid line and red dashed line
are the corresponding analytical results in the zero-range limit,
Eq. (\ref{eq:QAGPF}) and Eq. (\ref{eq:QAMF}). The blue dot-dashed
line shows the leading contribution to the deviation of the mode frequency:
$\delta\omega_{B}/(2\omega_{0})\simeq-\ln^{-1}(k_{F}R_{p})/8$; see,
Eq. (\ref{eq:QALeading}). }
\end{figure}

\subsection{Frequency shift}

The mean-field analytic expression and the GPF empirical formalism
for the chemical potential at resonance are very useful to understand
the shift of the breathing mode frequency. To see this, we may calculate
the polytropic index related to the chemical potential, i.e., $\mu\propto n_{2D}^{\gamma}$,
by using
\begin{equation}
\gamma=\frac{n_{2D}}{\mu}\left(\frac{\partial\mu}{\partial n_{2D}}\right)\simeq1+\frac{\xi_{y}^{(\mu)}}{2\xi^{(\mu)}},
\end{equation}
where we have rewritten $\mu=\varepsilon_{F}\xi^{(\mu)}(y)$ and have
assumed $\xi_{y}^{(\mu)}\ll\xi^{(\mu)}$. By taking the chemical potential
at the leading order, i.e., $\xi^{(\mu)}\propto-1/y$, we obtain immediately
$\gamma\simeq1-1/[2\ln(k_{F}R_{p})]$, and consequently,
\begin{equation}
\frac{\delta\omega_{B}}{\omega_{c}}\simeq\frac{\gamma-1}{4}\simeq-\frac{1}{8y}=-\frac{1}{8\ln\left(k_{F}R_{p}\right)}.\label{eq:QALeading}
\end{equation}
More careful treatments of the chemical potential to the next order
in Eqs. (\ref{eq:UnitaryMuMF}) and Eq. (\ref{eq:UnitaryMuGPF}) lead
to the results,
\begin{equation}
\frac{\delta\omega_{B}}{\omega_{c}}\simeq-\frac{1}{8y}\left[1-\frac{1}{2y}\ln\left(-\frac{e}{2y}\right)\right]\label{eq:QAMF}
\end{equation}
for the mean-field theory and
\begin{equation}
\frac{\delta\omega_{B}}{\omega_{c}}\simeq-\frac{1}{8y}\left[1-\frac{1}{2y}\right]\label{eq:QAGPF}
\end{equation}
for the GPF theory, respectively.

In Fig. \ref{fig7_unitarywb}, we show the up-shifts of the breathing
mode frequency predicted by the mean-field theory and the GPF theory
in the resonance limit, together with the asymptotic behaviors in
the zero-range limit, as discussed in the above. According to the
GPF theory, the shift of the breathing mode frequency can easily reach
10\% at a relatively small effective range, i.e., $[\ln(k_{F}R_{p})]^{-1}\simeq-0.6$
or $k_{F}R_{p}\sim0.2$.

At this point, it is interesting to compare the frequency shift exhibited
by a resonantly interacting $p$-wave Fermi superfluid and by a strongly
interacting $s$-wave Fermi superfluid, both in two dimensions. In
the latter case, the theoretically predicted maximum quantum anomaly
of 10\% is yet to be experimentally confirmed \cite{Vogt2012,Holten2018,Peppler2018}.
The main obstacle comes from the confinement-induced effective range
$R_{s}$, which is significant under the current experimental condition.
Indeed, our recent analysis indicates that the effective range $R_{s}$
in a 2D $s$-wave superfluid can strongly suppress the quantum anomaly
down to 1-2\% \cite{Hu2019}. In sharp contrast, for a resonantly
interacting $p$-wave Fermi superfluid, the effective range $R_{p}$
enhances the frequency shift. Owing to the great feasibility in tuning
$R_{p}$ in cold-atom experiment, therefore, we anticipate that a
low-temperature $p$-wave Fermi gas at Feshbach resonances would be
an ideal candidate to conclusively confirm the predicted frequency
shift.

\section{Conclusions and outlooks}

In conclusions, we have theoretically determined two important experimental
observables - Tan's contact parameter and the breathing mode frequency
- of a resonantly interacting $p$-wave Fermi superfluid in two dimensions
at the BEC-BCS evolution. Both observables can be easily accessed
in current cold-atom experiment, as soon as a stable $p$-wave superfluid
is realized in reduced dimensions. The two Tan's contact parameters
can be directly extracted from the tail of momentum distribution probed
by radio-frequency spectroscopy \cite{Luciuk2016}, and the breathing
mode measurement is now a routine tool in cold-atom laboratories \cite{Vogt2012,Holten2018,Peppler2018}.

We have proposed that, similar to an $s$-wave Fermi superfluid at
the BEC-BCS crossover, the $p$-wave Fermi superfluid in the resonance
limit experiences a frequency shift, due to the non-vanishing effective
range of interactions that explicitly breaks the scale invariance
of the system. The up-shift in the breathing mode frequency, away
from the scale-invariant value $\omega_{c}=2\omega_{0}$, turns out
to be significant. At the leading order, it is inversely proportional
to the logarithm of the effective range. As a result of this slow-decay
logarithmic dependence, the frequency shift can reach 5-10\% over
a wide range of the effective range.
\begin{acknowledgments}
We thank Shizhong Zhang and Yi-Cai Zhang for stimulating discussions.
This research was supported by Australian Research Council's (ARC)
Discovery Programs Grant No. DP170104008 (HH), Grant No. FT140100003
and Grant No. DP180102018 (XJL).
\end{acknowledgments}

\appendix

\section{An integral in the two-body $T$-matrix }

Here we consider the integral,
\begin{align}
\mathcal{I} & =\mathcal{\mathscr{P}}\sum_{{\bf p}}\frac{\left|\Gamma\left({\bf p}\right)\right|^{2}}{2\epsilon_{{\bf p}}-E},\nonumber \\
 & =\frac{M}{\hbar^{2}k_{F}^{2}}\intop_{0}^{\infty}\frac{pdp}{2\pi}\frac{p^{2}}{\left[1+\left(p/k_{0}\right)^{2n}\right]^{3}}\mathcal{\mathscr{P}}\frac{1}{p^{2}-k^{2}}.
\end{align}
By introducing the variable $z\equiv(p/k_{0})^{2}$, we find that,
\begin{equation}
\mathcal{I}=\frac{Mk_{0}^{2}}{4\pi\hbar^{2}k_{F}^{2}}\intop_{0}^{\infty}dz\frac{z}{\left(1+z^{n}\right)^{3}}\mathcal{\mathscr{P}}\frac{1}{z-z_{0}},
\end{equation}
where $z_{0}\equiv(k/k_{0})^{2}\ll1$. To handle the operator $\mathcal{\mathscr{P}}$
for Cauchy principle value, we divide the whole integral into three
parts $[0,z_{0})\cup[z_{0},2z_{0})\cup[2z_{0},\infty)$. Upon changing
the dummy variable, the integral $\mathcal{I}$ can be rewritten in
terms of $I_{1}$ and $I_{2}$,
\begin{equation}
\mathcal{I}=\frac{Mk_{0}^{2}}{4\pi\hbar^{2}k_{F}^{2}}\left[I_{1}+I_{2}\right],
\end{equation}
where
\begin{equation}
I_{1}=\intop_{0}^{z_{0}}\frac{dz}{z}\left\{ \frac{z_{0}+z}{\left[1+\left(z_{0}+z\right)^{n}\right]^{3}}-\frac{z_{0}-z}{\left[1+\left(z_{0}-z\right)^{n}\right]^{3}}\right\} 
\end{equation}
and
\begin{equation}
I_{2}=\intop_{z_{0}}^{\infty}\frac{dz}{z}\frac{z_{0}+z}{\left[1+\left(z_{0}+z\right)^{n}\right]^{3}}.
\end{equation}
It is clear that $I_{1}=2z_{0}+o(z_{0})$. For $I_{2}$, by neglecting
the higher contribution $o(z_{0})$, it can be separated into two
parts,
\begin{equation}
I_{2}=-2z_{0}+\intop_{0}^{\infty}\frac{dz}{\left(1+z^{n}\right)^{3}}+z_{0}\intop_{z_{0}}^{\infty}\frac{dz}{z\left(1+z^{n}\right)^{3}}.
\end{equation}
These two parts can be integrated out explicitly: 
\begin{eqnarray}
\intop_{0}^{\infty}\frac{dz}{\left(1+z^{n}\right)^{3}} & = & \frac{\pi\left(n-1/2\right)\left(n-1\right)}{n^{3}\sin\left(\pi/n\right)},\\
\intop_{z_{0}}^{\infty}\frac{dz}{z\left(1+z^{n}\right)^{3}} & = & -\ln z_{0}-\frac{3}{2n}+\frac{3z_{0}^{n}}{n}+o(z_{0}^{n}).
\end{eqnarray}
Putting $I_{1}$ and $I_{2}$ together, up to the order $o(z_{0})$
we obtain the expression,
\begin{equation}
\mathcal{I}=\frac{Mk_{0}^{2}}{4\pi\hbar^{2}k_{F}^{2}}\left[\frac{\pi\left(n-\frac{1}{2}\right)\left(n-1\right)}{n^{3}\sin\left(\pi/n\right)}-z_{0}\ln z_{0}-\frac{3z_{0}}{2n}\right].
\end{equation}
which is Eq. (\ref{eq:INT2BTMatrix}) in the main text.

\section{Quantum anomaly in a strongly interacting $s$-wave Fermi superfluid}

In an $s$-wave Fermi superfluid, Tan's adiabatic relation is given
by \cite{Werner2012},
\begin{equation}
\left(\frac{\partial E}{\partial\ln a_{2D}}\right)_{S}=\frac{\hbar^{2}}{2\pi M}C,\label{eq:AdiabaticRelationContact}
\end{equation}
which takes exactly the same form as the adiabatic relation for the
effective range of interactions, as given in Eq. (\ref{eq:AdiabaticRelationCR}),
up to an unimportant pre-factor. This same form emphasizes the similar
role played by the effective range $R_{p}$ in a $p$-wave Fermi superfluid
and by the scattering length $a_{2D}$ in an $s$-wave Fermi superfluid.

Let us now write the total energy of the $s$-wave Fermi superfluid
in a dimensionless form \cite{He2015},
\begin{equation}
E=\frac{N\varepsilon_{F}}{2}\xi\left[z=\ln\left(k_{F}a_{2D}\right)\right],
\end{equation}
where for the two-component Fermi gas the Fermi wavevector $k_{F}=\sqrt{2\pi n_{2D}}.$
By using the adiabatic relation Eq. (\ref{eq:AdiabaticRelationContact}),
we then find,
\begin{equation}
\frac{C}{k_{F}^{4}}=\frac{1}{4}\xi_{z},
\end{equation}
where $\xi_{z}=\partial\xi/\partial z$. The dimensionless chemical
potential and pressure are also easy to obtain,
\begin{eqnarray}
\mu & = & \varepsilon_{F}\left(\xi+\frac{1}{4}\xi_{z}\right),\\
P & = & P_{0}\left(\xi+\frac{1}{2}\xi_{z}\right).
\end{eqnarray}
By calculating the polytropic index related to the pressure \cite{Hofmann2012},
we find that,
\begin{equation}
\gamma=\frac{n_{2D}}{P}\left(\frac{\partial P}{\partial n_{2D}}\right)-1\simeq1+\frac{\bar{\xi}_{z}}{2\bar{\xi}},
\end{equation}
where again the bar denotes the exclusion of the two-body bound-state
contribution. The calculation of the polytropic index related to the
chemical potential leads to the same expression at the same level
of approximation. Thus, we obtain the quantum anomaly,
\begin{equation}
\frac{\delta\omega_{B}}{2\omega_{0}}\simeq\frac{\gamma-1}{4}\simeq\frac{\bar{\xi}_{z}}{8\bar{\xi}}=\frac{C_{MB}/k_{F}^{4}}{2(E+N\varepsilon_{B}/2)/E_{0}}.
\end{equation}
According to the GPF calculation or quantum Monte Carlo simulations,
at around the strongly interacting regime $\ln\left(k_{F}a_{2D}\right)\sim0$,
the many-body part of the contact shows a peak with $C_{MB}/k_{F}^{4}\sim0.05$
\cite{He2015}. This is correlated with a total energy $(E+N\varepsilon_{B}/2)/E_{0}\sim0.25$
\cite{He2015}. By using these two numbers, we find a quantum anomaly
$\delta\omega_{B}/(2\omega_{0})\sim0.1$ for a strongly interacting
$s$-wave Fermi superfluid.


\begin{thebibliography}{10}
\bibitem{NoteVirialTheorem}There is a subtlety here, as we want to
relate the peak value of a quantity at the trap center to an average
of the quantity over the whole trap. In a 2D harmonic trap, the virial
theorem states that the trapping potential energy $\left\langle M\omega_{0}^{2}\sum r_{i}^{2}/2\right\rangle =M\omega_{0}^{2}\left\langle \mathcal{O}\right\rangle /2=\int d\mathbf{r}P(\mathbf{r})\simeq PA$.
In the last step, we approximate the integral using the peak pressure
$P$ at the trap center and define an appropriate area $A$, which
is to be removed by using the identity $P_{0}A=N\varepsilon_{F}/2$.
\end{thebibliography}
\end{document}